\documentclass[twocolumn,showpacs,amsmath,amssymb,prl,superscriptaddress]{revtex4}

\usepackage{graphicx}
\usepackage{bm}
\usepackage{color}

\begin{document}

\title{Low-voltage current noise in long quantum SINIS junctions}
\author{N. B. Kopnin}
\affiliation{Low Temperature Laboratory, Helsinki University of
Technology, P.O. Box 2200, FIN-02015 HUT, Finland and\\
L.D. Landau Institute for Theoretical Physics, 117940 Moscow,
Russia}
\author{Y. M. Galperin}
\affiliation{Department of Physics \& Center for Advanced
Materials and Nanotechnology, University of Oslo, PO Box 1048
Blindern, 0316 Oslo, Norway and \\ A. F. Ioffe Physico-Technical
Institute of Russian Academy of Sciences, 194021 St. Petersburg,
Russia}
 \affiliation{Argonne National Laboratory, 9700 S. Cass
Av., Argonne, IL 60439, USA}
\author{V. Vinokur}
\affiliation{Argonne National Laboratory, 9700 S. Cass Av.,
Argonne,
  IL 60439, USA}
\date{\today}

\begin{abstract}
The current noise in long SINIS junctions at low temperatures is
sensitive to the population of the sub-gap states which is far
from equilibrium even at low bias voltages. Nonequilibrium
distribution establishes due to an interplay between
voltage-driven inter-level Landau-Zener transitions and
intra-level inelastic relaxation. We show that the Fano factor is
enhanced drastically, being proportional to the number of times
which particle can fly along the Andreev trajectory before it
escapes from the level due to inelastic scattering. Combining the
dc current and noise measurements one can fully characterize the
non-equilibrium kinetics in SINIS junctions.
\end{abstract}


\maketitle

Charge transport through mesoscopic normal-superconductor (NS)
hybrid structures is a subject of intense current interest (see
\cite{Beenakker-rev,currentRev} for a review).  The multiple
Andreev reflections (MAR) at the NS interfaces lead to several
unusual features, in particular to the so-called sub-gap structure
in the current-voltage (I-V) curves~\cite{theory,KMV06,BTK}.  The
electric current noise in NS hybrid systems provides information
complementary to the dc-studies and exhibits properties that are
essentially different from those of the contacts between the
normal metals. In normal contacts, the shot noise induced by the
Pauli correlation is relatively weak ~\cite{correlation} and
vanishes if the interface is ideal. In NS systems, on the
contrary, Andreev scattering enhances reflection of electrons and
results in a finite shot noise even when the contact itself is
noiseless in the normal state~\cite{Hessling96}.  The situation at
low voltages is especially interesting: the interplay between the
quasiparticle scattering and the superconducting coherence reveals
an intricate structure of the non-equilibrium population of
sub-gap states and leads to considerable enhancement of the
noise~\cite{AverinImam96,Martin-Rod96,NavehAverin99,expt2,expt1}.

In the present work we study the noise in a single-mode long SINIS
junction (I stands for an insulator) where the length $d$ of the
normal conductor is much larger than the superconducting coherence
length, but shorter than the inelastic mean-free path. This is an
exemplary hybrid structure~\cite{nanotubes} where non-equilibrium
dynamics is governed by inelastic scattering and voltage-driven
Landau-Zener (LZ) transitions between Andreev sub-gap states. In
particular, in a long junction where the number of the states
involved in the transport is large, the LZ transitions lead to a
significant increase in the dc current far above the critical
Josephson current of the junction~\cite{KV07}. The noise
properties of such systems are very informative. At finite
temperature, the current noise is finite even in the absence of a
dc bias voltage (and dissipation) and is related to thermal
fluctuations in the occupancies of the sub-gap
states~\cite{AverinImam96,Martin-Rod96,NavehAverin99}. The width
of the zero-frequency peak in the noise spectrum is determined by
the inelastic relaxation rate and thus provides an experimental
tool for measuring its temperature dependence. We show that for
low temperatures and small bias voltages the noise is controlled
by the non-equilibrium population of the sub-gap states
characterized by an {\em effective temperature} $T_{\rm eff}$
which is determined by the competition between the inelastic
scattering and the voltage-driven inter-level LZ transitions;
$T_{\rm eff}$ depends on the bias voltage and remains finite even
at zero ambient temperature. At the same time, the dc current
depends not only on $T_{\rm eff}$ but also on a LZ-controlled
parameter that accounts for the charge transfer from the excited
levels. For weak LZ processes, this parameter and thus the dc
current are small. As a result, the ratio of the zero-frequency
noise to the dc current, i. e., the Fano factor proportional to
the fluctuating ``effective charge'' is increased drastically by a
factor qualitatively different from that predicted in
Refs.~\cite{AverinImam96,NavehAverin99} for short contacts. Since
both the dc I-V curve and the noise spectrum depend on inelastic
relaxation rate and on the LZ probabilities, the combined I-V and
noise measurements enable detailed characterization of
non-equilibrium dynamics in long SINIS junctions.

\paragraph{Noise spectrum.}
The current operator is
\begin{equation}
  \label{eq:001}
 \hat I(t_1,x_1)=\frac{ie\hbar}{2m}\left[ \left(\nabla_1 -\nabla
^\prime _1\right) \sum_\alpha \hat \psi^\dagger_\alpha (1)\hat
\psi_\alpha (1^\prime )\right]_{1=1^\prime}
\end{equation}
where $\hat \psi_\alpha$ is the operator of particle field with a
spin $\alpha$. Noise spectrum is defined as a Fourier transform,
\begin{equation}
\label{eq:002}
S(\omega) =\int \! \! e^{i\omega \tau} d\tau
\overline{ \left \langle \! \left\{\Delta \hat
I(t-\frac{\tau}{2},x),\Delta \hat
  I(t+\frac{\tau}{2},x)\right\}\!\right \rangle}
\end{equation}
where $\Delta \hat I \equiv \hat I - \langle \hat I\rangle$,
$\langle \hat I \rangle$ is the statistical average current, and
$\{\hat A,\hat B\} \equiv (\hat A \hat B +\hat B \hat A)/2$, while
overline means the average over a long time interval, $T_0$:
$\overline{ A} \equiv T_0^{-1}\int_0^{T_0} A(t)\,  dt$.

We focus on the low-frequency behavior of the noise correlator,
such that $\hbar \omega $ is much smaller than the characteristic
energies of the states involved. In this situation, making use of
the Bogoliubov transformation one can express the correlator in
Eq.\ (\ref{eq:002}) through the occupation numbers $f(\epsilon_n)$
of the sub-gap states and matrix elements, $L_{mn}$, quadratic in
current operators:
\begin{eqnarray}
&&\hspace*{-2.2mm} S= -\frac{\pi \hbar^3e^2}{4m^2}
\overline{\sum_{n,m}\! \! \delta(\epsilon_{nm} +\hbar \omega)
L_{nm} \! \left[ 1- f_1(\epsilon_n) f_1(\epsilon_m)\right]}\ ,
\label{eq:Somega} \\
&& \!\!\! L_{nm}(x,t)=\left\{ \left(\nabla_1 -\nabla
    ^\prime_1\right)\left(\nabla_2 -\nabla ^\prime _2\right)
\left[ u^*_n(1)u_m(1^\prime ) \right. \right.  \\ && \!\!\!
\left.\left. + v_n^*(1)v_m(1^\prime)\right] \left[ u_n(2^\prime
)u_m^*(2)+v_n(2^\prime)v_m^*(2)\right] \right\}_{1=1^\prime
=2=2^\prime }.  \nonumber \label{L-nm}
\end{eqnarray}
Here $u_n,v_n$ and $\epsilon_n$ are the wave functions and
energies of the sub-gap states, which have to be found from the
Bogoliubov-de Gennes (BdG) equations,
$\epsilon_{nm}=\epsilon_n-\epsilon_m$. These quantities depend
adiabatically on time $t$. The matrix elements $L_{nm}$ are
symmetric in the transposition $n\leftrightarrow m$. The functions
$f(\epsilon_n)$ are the occupation numbers of the subgap states;
here we have introduced the odd-in-$\epsilon$ combination
$f_1(\epsilon)\equiv f(-\epsilon)-f(\epsilon)$. We will see later
that the even combination does not appear in our case, $f_2\equiv
1-f(\epsilon)-f(-\epsilon) =0$.

\paragraph{Sub-gap states.}
To find the matrix elements $L_{mn}$ we solve the BdG equations
for $\epsilon <|\Delta|$ modelling the SINIS structure as two
insulating barriers described  by a symmetric $\delta$-function
potential with a magnitude $I$. By adjusting the barrier height
$I$ one can account both for the velocity mismatch in the adjacent
conductors and for the tunneling of electrons through the
interface. The gap $\Delta =|\Delta|e^{i\chi_{R,L}}$ in the right,
$x>d/2$, and left, $x<-d/2$, superconductors has $\chi_{R,L}=\pm
\phi/2$, respectively, while $\Delta =0$ in the normal conductor,
$-d/2<x<d/2$. The states depend on the phase parameters $\alpha =
k_xd$ and  $\beta =\epsilon d/\hbar v_x$, as well as on the
unitary scattering matrices\cite{Beenakker-rev} $\hat S$ at each
NIS interface
\begin{equation}
\hat S=
\left( \begin{array}{cc} S_{N}e^{i\delta} & S_{A}e^{i\chi} \\
S_{A}e^{-i\chi} & S_{N}e^{-i\delta}\end{array}\right)
\end{equation}
which is parameterized through the barrier strength $Z\equiv
mI/\hbar ^2k_x$ and the scattering phase $\delta =\arctan 1/Z$.
Here $k_x$ and $v_x$ are the electron wave vector and velocity,
respectively. See Refs.~\cite{KMV06,BTK} for more details and
expressions for $S_N$, $S_A$.

The wave function inside the normal conductor is a linear
combination of an incident particle and an incident hole, each
experiencing Andreev and normal reflections at the interfaces. For
reflections from the right barrier at $x=d/2$ we have
\begin{eqnarray}
 \left( \begin{array}{c} u  \\ v
\end{array}\right)&=&b\left[\hat u^>
+S_Ae^{-i\phi /2}\hat v^< +S_Ne^{i\delta}\hat u^<\right]\nonumber \\
&& +a\left[\hat v^>+S_Ae^{i\phi /2}\hat u^< +S_Ne^{-i\delta}\hat
v^<\right];
\label{function1} \\
\hat u^{>,<}(x)&=&e^{\pm iq_+x}\left( \begin{array}{c} 1  \\
0\end{array}\right), \; \hat v^{<,>}(x)=e^{\pm iq_-x}\left(
\begin{array}{c} 0  \\ 1\end{array}\right),
\nonumber
\end{eqnarray}
where $q_\pm = k_x\pm \epsilon/\hbar v_x$. Comparing Eq.
(\ref{function1}) with the similar equation for the left barrier
at $x=-d/2$ we find
\begin{eqnarray}
ae^{i\alpha}\left[ e^{-2i\beta} -e^{i\phi}S_A^2
-e^{-2i\alpha^\prime }S_N^2\right]&& \nonumber \\
-2 bS_AS_N\cos (\alpha^\prime +\phi /2)&=&0\ , \label{eq-scat1}\\
be^{-i\alpha}\left[ e^{-2i\beta} -e^{2i\alpha^\prime }S_N^2
-e^{-i\phi /2}S_A^2\right]&& \nonumber \\
- 2aS_AS_N\cos (\alpha^\prime +\phi /2)&=&0\ . \label{eq-scat2}
\end{eqnarray}
Here $\alpha^\prime = \alpha +\delta $. The wave function Eq.\
(\ref{function1}) is normalized $\int (|u|^2+|v|^2)\, dx =1$.
Defining $ S_N=e^{i\gamma}|S_N| $, the solvability of
Eqs.~(\ref{eq-scat1})--(\ref{eq-scat2}) requires~\cite{KMV06}
\begin{equation}
|S_{N}|^2\sin ^2 \alpha ^\prime +|S_{A}|^2\cos ^2(\phi /2)
=\sin^2(\beta +\gamma)\ . \label{determinant2}
\end{equation}
A long contact has $d\gg \hbar v_x /\Delta$ and a large number of
levels $N\sim d\Delta/\hbar v_x\gg 1$. The levels are separated by
minigaps at $\phi = 2\pi k$ and $\phi = \pi(2k+1)$ which disappear
when $\alpha^\prime =\pi /2+\pi m$ and $\alpha^\prime =\pi m$,
respectively.

\paragraph{Equilibrium noise.}

For frequencies $\hbar \omega$ much lower than the interlevel
spacing  $ \epsilon_{n,n+1}\sim \hbar v_x/d$, only diagonal
elements contribute to Eq.\ (\ref{eq:Somega}) due to the $\delta$
functions,
\begin{eqnarray*}
L_{nn}&=&-16 k_x^2\left[ (|a_n|^2-|b_n|^2)|S_{A}|^2
\right.  \\
&&\left. + a_n^*b_n S_{N}S_{A}^*e^{i\delta -i\phi /2} + a_nb_n^*
S_{N}^*S_{A} e^{-i\delta +i\phi /2}\right]^2\ .
\end{eqnarray*}
Using Eqs.~(\ref{function1})--(\ref{determinant2}) for the cases
of nearly ballistic and weakly transparent contacts we obtain,
respectively,
\begin{equation} \label{L}
\hspace*{-0.05in}
  L_{nn}=-\frac{4k_x^2}{d^2}\left\{
\begin{array}{ll}
1\ ,& Z\to 0, \ |S_N|\to 0;\\ |S_A|^4\begin{displaystyle} \frac{
\sin^2\phi}{\sin^2 \alpha'} \end{displaystyle}\ , & Z\to \infty,
|S_A|\to 0\, .
\end{array}
\right.
\end{equation}

Putting $f_1(\epsilon)=f_1 ^{(0)}(\epsilon)+\tilde f_1(\epsilon)$
where $f_1 ^{(0)}(\epsilon)=\tanh(\epsilon/2T)$ is the equilibrium
distribution we find
\begin{equation}
\hspace*{-0.05in}
 S(\omega)= -\frac{\pi \hbar
^2e^2}{8m^2}\delta(\omega)\!  \sum_{n}  \overline{L_{nn} \left[
1-[f_{1}^{(0)}]^2 -2f_{1}^{(0)}\tilde f_{1} \right]}\! .
\label{S-lowomega}
\end{equation}

The first two terms in the square brackets in Eq.
(\ref{S-lowomega}) give the equilibrium thermodynamic noise with
$1-[f_{1}^{(0)}(\epsilon)]^2=\cosh ^{-2}(\epsilon/2T)$. The
$\delta$-function in Eq.\ (\ref{S-lowomega}) is broadened by
inelastic relaxation processes described by the rate $\tau^{-1}$
corresponding to energies $\epsilon \ll \Delta$. Since the
interlevel spacing $\epsilon_{n,n+1}\gg \hbar/\tau$ while $\hbar
\omega \ll \epsilon_{n,n+1}$ the $\delta(\omega)$ should be
replaced with
\[
\Lambda (\omega) = \frac{1}{2\pi}\, \frac{\tau^{-1}}{\omega ^2
+1/4\tau^2}\ .
\]

For high transparency we get from
Eqs.~(\ref{L})-(\ref{S-lowomega})
\begin{equation}
S_{\rm th}(\omega)=\frac{\pi v_x^2e^2}{d^2}
\Lambda(\omega)\overline{ \sum_{n>0}
\cosh^{-2}\frac{\epsilon_n}{2T}} \ .
\end{equation}
If $\hbar v_x/d \ll T \ll \Delta$, the energies $\epsilon \ll
\Delta$. Replacing the sum with the integral $ (2d/\pi \hbar
v_x)\int d \epsilon $ we find
\begin{equation}
S_{\rm th}(\omega) = (4T v_x e^2/ \hbar d) \Lambda(\omega)\ .
\label{S-omega-therm}
\end{equation}
For a low-transparency junction, $Z\rightarrow \infty$, one has
$|S_A|^2 =1/4Z^4$ for $\epsilon \ll \Delta$, and the noise
acquires an additional factor $(32Z^8\sin^2 2\alpha')^{-1}$ as
compared to Eq.~(\ref{S-omega-therm}). The current noise in Eq.\
(\ref{S-omega-therm}) is associated with thermal fluctuations in
the occupancies of  the sub-gap states in the same manner as
obtained in Ref.~\cite{AverinImam96} for short junctions. However,
the spectral density in long junctions is linear in temperature
which accounts for the number of excited levels $N_{\rm eff}=
2Td/\pi \hbar v_x$ contributing to the noise.

\paragraph{Non-equilibrium distribution in high-transparency junctions.}

A dc current at a finite bias voltage appears only due to
nonequilibrium quasiparticle distribution. Indeed, in equilibrium
the current would be just the supercurrent adiabatically driven by
the time-dependent phase difference $\phi = \omega_J t+ \phi_0$
and having zero time-average ($\omega_J=2eV/\hbar $ is the
Josephson frequency). In a highly-transparent barrier, particles
with $|\epsilon|>|\Delta|$ remain in equilibrium with the heat
bath for $eV \ll \Delta$ because they have sufficient time to
escape from the double-barrier region and relax in the continuum.
However, the time depending phase induces inter-level LZ
transitions across the minigaps near $\phi = \pi k$ producing
deviation from equilibrium for the sub-gap states. As the phase
changes with time, the outermost Andreev levels split off from the
continuum and capture particles at energies $\epsilon =\pm
|\Delta|$. These quasiparticles can be then excited to the other
levels by LZ transitions. The interplay between the LZ transitions
and the weak inelastic relaxation produces the nonequilibrium
population of the subgap states and leads to a finite dc current.

The kinetic equation for the subgap distribution was recently
derived in Ref.~\cite{KV07} under assumption that LZ transitions
occur only near the avoided crossings at $\phi = \pi k$. This can
be realized if the minigaps as well the bias, $eV$, are much less
than the typical distance $\hbar v_x/d$ between the levels. Only
in this case the avoided crossing points are well defined and
there are no voltage-induced inter-level transitions elsewhere.
These requirements can be met only in almost ballistic contacts,
$1-{\cal T}\ll 1$, where $\cal T$ is the transparency of the
interface. The kinetic equation also assumes incoherent LZ
transitions. Indeed, the inequality $eV \ll \hbar v_x/d$ ensures
that the correlation time of the quantum beating between the
adjacent Andreev states, $\hbar/|\epsilon_{n+1,n}| \approx d/v_x
$, is much shorter than the time $\sim \omega_J^{-1}=\hbar/2eV$
between successive level crossings.

Between the LZ tunnelling events, the distribution function
relaxes due to inelastic scattering with the rate $\tau^{-1}$,
which is assumed to be energy-independent. Since the even
component is absent, $f_2=0$, we have $f_1(\epsilon_n)\equiv
1-2f(\epsilon_n)$. For low temperatures, $T\ll \hbar v_x/d$, the
equilibrium distribution $f^{(0)}_1(\epsilon_n) =\mathrm {sign}\,
(\epsilon_n) $. As a result,
\[
f_1[\epsilon_n(\phi)]-f_1^{(0)}(\epsilon_n)=\left\{
\begin{array}{lr}\psi_n e^{-\hbar \phi /2eV\tau} ,
& 0 <\phi < \pi, \\
\chi_n e^{-\hbar (\phi - \pi)/2eV\tau}  , & \pi <\phi < 2\pi.
\end{array} \right.
\]

The LZ transitions at $\phi =2\pi k$ and $\phi =\pi +2\pi k$ with
(constant) probabilities $p_0$ and $p_\pi$, respectively, lead to
difference equations for $\psi_n $ and $\chi_n $ with constant
coefficients\cite{KV07}. The boundary conditions require that, for
the levels moving upwards with the increase in $\phi$, the
distribution coincides with the equilibrium at $\epsilon
=-|\Delta|$, while, for levels moving downwards, the distribution
coincides with the equilibrium at $\epsilon =+|\Delta|$. Solution
to the difference equations~\cite{KV07} obeys $\psi_{-n} =-\psi
_{n+1}$, which implies the property
$f_1(\epsilon)=-f_1(-\epsilon)$ used earlier. For $\epsilon_n>0$
we have $ \psi_{n+1} +\psi_{n} =-2e^{\nu}F(N-n)/F(N) $, where $
\nu = \pi \hbar /2eV\tau $ is the degree of inelastic relaxation
during the Josephson half-period,
\begin{eqnarray*}
F(z)&=&e^{rz}\left(1+e^{r}w_+\right)-
e^{-rz}\left(1+e^{-r}w_-\right)\, , \\
w_\pm &=& \frac{p_0e^{\pm r}+p_\pi e^{\mp r}-2p_0p_\pi \cosh
(r)}{\zeta +p_0p_\pi (1-e^{\pm 2r})+p_0+p_\pi -2p_0p_\pi} \, ,
\end{eqnarray*}
and $\zeta =e^{2\nu} -1 $. Depending on voltage, $\nu$ can be
either large or small. For very low voltage, $\nu \gg 1$, the
relaxation rate is $r\approx \nu$, and the distribution relaxes
quickly. The most interesting case is when the degree of inelastic
relaxation is small, $\nu \ll 1$. In this limit, $r$ is given by
\begin{equation}
\sinh^2 (r)=\nu (\nu +p_0+p_\pi- 2p_0p_\pi)/p_0p_\pi \ .
\label{r-expr}
\end{equation}
The inverse rate $r^{-1}$ describes the energy broadening of the
distribution and determines an effective temperature
\begin{equation}
T_{\text {eff}}= r^{-1}(d\epsilon_n /dn)=(\pi \hbar v_x/2rd)
\label{Teff}
\end{equation}
which can much exceed the interlevel spacing if $r\ll 1$.

Let $N=2d\Delta/\pi\hbar v_x$ be the number of sub-gap levels with
$\epsilon_n>0$. For cases (\textit{i}) $1/N \ll r\ll 1$ and
(\textit{ii}) $r \ll 1/N$ we have
\begin{equation} \label{sum-psi}
\psi_{n+1} +\psi_{n} = -2 \left\{\!\! \begin{array}{ll} e^{-rn}, &
(i);
\\ \begin{displaystyle}
 \frac{N+(p_0+p_\pi
-2p_0p_\pi)^{-1} -n}{N+(p_0+p_\pi
-2p_0p_\pi)^{-1}}
\end{displaystyle}
, & (ii). \end{array} \right.
\end{equation}

\paragraph{Nonequilibrium noise.}

The equilibrium noise vanishes if $T\ll \hbar v_x/d$. For high
transparency $Z\rightarrow 0$ and slow relaxation $\nu \ll 1$ the
nonequilibrium part of Eq. (\ref{S-lowomega}) is
\[
S_{\rm ne}(\omega)= -\frac{2\pi e^2v_x^2 }{d^2} \Lambda(\omega)
\sum_{k=0}^{N/2} \left[\psi_{n+1} +\psi_{n}\right]_{n=2k+1}.
\]
Using Eq.~(\ref{sum-psi})  one finds
\begin{eqnarray}
S_{\rm ne}(\omega)&=&(2\pi e^2 v_x^2 N_{\rm eff} /d^2)
\Lambda(\omega)\, ;
\label{Somega-N} \\
N_{\text{eff}}&=&\left\{ \begin{array}{lr} r^{-1} \ , & (i)\, ; \\
\begin{displaystyle}\frac{N}{2}\, \frac{2[N(p_0+p_\pi
-2p_0p_\pi)]^{-1}+1}{[N(p_0+p_\pi -2p_0p_\pi)]^{-1}+1}
\end{displaystyle} \ , &(ii)\, . \end{array} \right. \nonumber
\end{eqnarray}
$N_{\text{eff}}$  is the number of excited levels. In the case (i)
the noise $S_{\rm ne}(\omega)$ has the form of
Eq.~(\ref{S-omega-therm}) where the real temperature is replaced
with the {\em effective temperature} $T_{\text{eff}}=\pi \hbar
v_xN_{\rm eff}/2d$ defined by Eq.~(\ref{Teff}). This result agrees
qualitatively with the low-voltage noise in short contacts
obtained in Ref. \cite{AverinImam96} if the real temperature there
is replaced with $T_{\text{eff}}$. The zero-frequency noise is
\begin{equation}
S_{\rm ne}(0)=(4e^2 v_x^2 /d^2)N_{\text{eff}}\tau= (8T_{\rm
eff}e^2 v_x \tau / \pi \hbar d) \ . \label{Szero}
\end{equation}
This noise does not vanish at zero real temperature; on the
contrary, it diverges for $\tau \rightarrow \infty$ since $N_{\rm
eff}\sim N$ and $T_{\rm eff}\sim \Delta$ when $\nu \rightarrow 0$.
The effective temperature depends on voltage; it decreases
exponentially for low voltages and grows linearly for higher
voltages (see below).

The dc current calculated in Ref.~\cite{KV07} is
\begin{equation}
I=(ev_x/d)\times \left\{ \begin{array}{lr} N_{\rm eff}\lambda \ , & (i);\\
N \ ,  & (ii). \end{array} \right.  \ \label{I-N}
\end{equation}
The factor $\lambda\equiv  \nu /r =N_{\text{eff}}\nu$ is found
from Eq.\ (\ref{r-expr})
\[
\lambda = \sqrt{\frac{p_0p_\pi}{1+(p_0+p_\pi-2p_0p_\pi)/\nu}} \leq
1 \ .
\]
It is a LZ-controlled efficiency of the non-equilibrium
distribution in contributing to the charge transfer through the
junction; $\lambda =1$ if $p_0=p_\pi =1$, while $\lambda =0$ if
$p_0=p_\pi =0$: there is no dc current if $\lambda \rightarrow 0$.

One can write $ S_{\rm ne}(0)= 2q I=2eIF $ where $F$ is the Fano
factor, and $q=eF$ is the effective charge $ q=2en /\lambda $.
Here $n=v_x \tau/d$ is the number of times which a particle can
fly back and forth along the Andreev level trajectory before it
escapes due to inelastic scattering. The effective charge is huge
$q\gg e$ since $\tau \gg d/v_x$. The enhancement of fluctuating
charge for MAR in short junctions has been predicted theoretically
in Refs.~\cite{AverinImam96,NavehAverin99} and confirmed
experimentally in Refs.~\cite{expt2,expt1}. For a finite bias
voltage the charge is enhanced by a factor $n= \Delta/eV$ which is
the number of flights of a particle in the normal region before it
accumulates the energy sufficient to escape into continuum. Here
we predict that, in long junctions, the fluctuating charge is
increased  by a larger factor $n/\lambda=v_x \tau/d\lambda$ for
bias voltages $eV \ll \hbar v_x/d $.

For highly-transparent contacts $1-{\cal T} \ll 1$ and low
energies, the probabilities $p_0$ and $p_\pi$ of LZ tunneling  can
be easily calculated. Assuming that $ \phi = \omega _Jt + \phi_0 $
one finds \cite{KV07} $ p_\pi =\exp \left[ -(\omega _0/\omega
_J)\sin^2\alpha^\prime \right] $ where $ \omega_0=\pi v_x(1-{\cal
T}^2)/{\cal T}d $. For $p_0$ one replaces $\sin \alpha^\prime$
with $\cos \alpha^\prime$. Therefore, for $\omega_J\ll \omega_0$
the efficiency factor $\lambda$ is small, leading to enormous
enhancement of the effective charge. For $\omega_J\gg \omega_0$
one obtains $ p_0+p_\pi-2p_0p_\pi =\omega_0/\omega_J $. In the
case $Nr\gg 1$ the effective temperature is linear in voltage,
$T_{\text{eff}}=(\hbar v_x/2d)(\omega _J \tau/ \sqrt{1+\omega _0
\tau /\pi})$. The LZ efficiency $
\lambda=1/\sqrt{1+\omega_0\tau/\pi} $ is much smaller than unity
for $\omega _0\tau \gg 1$. As a result, the effective charge is
still much larger than $2en$, namely
$q=2en\sqrt{1+\omega_0\tau/\pi} $.

In conclusion, we have considered both thermal and non-equilibrium
low-voltage noise spectrum in long SINIS junctions. The spectrum
has a pronounced peak at $\omega=0$; its intensity diverges at
zero temperature, while the width is equal to the inelastic
relaxation rate. The spectrum allows one to determine the
effective temperature of the sub-gap excitations which remains
finite even at zero ambient temperature. At weak inelastic
scattering the Fano factor is large; it is determined by LZ
tunnelling between the sub-gap states.

\acknowledgments This work was partly supported by the Academy of
Finland (grant 213496, Finnish Programme for Centres of Excellence
in Research 2002-2007/2006-2011), by the ULTI program under EU
contract RITA-CT-2003-505313, by the U.S. Department of Energy
Office of Science contract No. DE-AC02-06CH11357, and by the
Russian Foundation for Basic Research grant 06-02-16002. We are
grateful to J. Bergli for helpful comments.

\end{document}